%% file: main.tex
\documentclass[11pt]{article}
\input{setup.tex}

\begin{document}

\title{Variational Quantum Evolution Equation Solver}
\author[]{Fong Yew Leong\orcidlink{0000-0002-0064-0118}\thanks{Corresponding author, leongfy@ihpc.a-star.edu.sg} }
\author[]{Wei-Bin Ewe\orcidlink{0000-0002-4600-0634}\thanks{ewewb@ihpc.a-star.edu.sg} }
\author[]{Dax Enshan Koh\orcidlink{0000-0002-8968-591X}\thanks{ dax\textunderscore koh@ihpc.a-star.edu.sg} }
\affil[]{\small Institute of High Performance Computing, Agency for Science, Technology and Research (A*STAR), \\
1 Fusionopolis Way, \#16-16 Connexis, Singapore 138632, Singapore}
\date{}

\maketitle

\begin{abstract}
Variational quantum algorithms offer a promising new paradigm for solving partial differential equations on near-term quantum computers. Here, we propose a variational quantum algorithm for solving a general evolution equation through implicit time-stepping of the Laplacian operator. The use of encoded source states informed by preceding solution vectors results in faster convergence compared to random re-initialization. Through statevector simulations of the heat equation, we demonstrate how the time complexity of our algorithm scales with the ansatz volume for gradient estimation and how the time-to-solution scales with the diffusion parameter. Our proposed algorithm extends economically to higher-order time-stepping schemes, such as the Crank-Nicolson method. We present a semi-implicit scheme for solving systems of evolution equations with non-linear terms, such as the reaction-diffusion and the incompressible Navier-Stokes equations, and demonstrate its validity by proof-of-concept results.
\end{abstract}


\section{Introduction}

Partial differential equations (PDEs) are fundamental to solving important problems in disciplines ranging from heat and mass transfer, fluid dynamics and electromagnetics to quantitative finance and human behavior. Finding new methods to solve PDEs more efficiently---including making use of new algorithms or new types of hardware---has been an active area of research.

Recently, the advent of quantum computers and the invention of new quantum algorithms have provided a novel paradigm for solving PDEs. A cornerstone of many of these quantum algorithms is the seminal Harrow-Hassidim-Lloyd (HHL) algorithm \cite{harrow2009quantum} for solving linear systems, which can be utilized to solve PDEs by discretizing the PDE and mapping it to a system of linear equations. Compared to classical algorithms, the HHL algorithm can be shown to exhibit an exponential speedup. Unfortunately, attractive as it may sound, the HHL algorithm works only in an idealized setting, and a list of caveats must be addressed before it can be used to realize a quantum advantage \cite{aaronson2015read}. Moreover, implementing HHL and many other quantum algorithms would require the use of a fault-tolerant quantum computer, which may not be available in the near future \cite{bharti2022noisy}. Instead, the machines we have today are imperfect, noisy intermediate-scale quantum (NISQ) devices \cite{preskill2018quantum} with both coherent and incoherent errors limiting practical circuit depths.

Over the last few years, variational quantum algorithms (VQAs) have emerged as a leading strategy to realize a quantum advantage on NISQ devices. Specifically, VQAs employ shallow circuit depths to optimize a cost function, expressed in terms of an ansatz with tunable parameters, through iterative evaluations of expectation values \cite{cerezo2021variational}. Applications of VQAs include the variational quantum eigensolver (VQE) for finding the ground or excited states of a system Hamiltonian \cite{peruzzo2014variational,mcclean2016theory,kandala2017hardware}, the quantum approximate optimization algorithm (QAOA) for solving combinatorial optimization problems \cite{farhi2014quantum}, and solvers for linear \cite{bravoprieto2020variational,huang2019near,Xu2019variational} and non-linear \cite{lubasch2020variational} systems of equations. 

Here, we are interested in variational quantum algorithms for solving differential equations \cite{arrazola2019quantum}, such as the Black-Scholes equation \cite{fontanela2021quantum,miyamoto2021pricing}, the Poisson equation \cite{liu2021variational,sato2021variational}, and the Helmholtz equation \cite{ewe2021variational}. Specifically, the Poisson equation can be solved efficiently through explicit decomposition of the coefficient matrix derived from finite difference discretization \cite{liu2021variational} using minimal cost function evaluations \cite{sato2021variational} and shallower circuit depth compared to other non-variational quantum algorithms \cite{cao2013poisson,linden2020quantum,childs2021highprecision,arrazola2019quantum}. A natural question to ask, then, is whether such variational algorithms for Poisson equations can be extended to solving evolution equations, i.e.~partial differential equations including a time domain. McArdle et al.~\cite{mcArdle2019variational} proposed a variational quantum algorithm which simulates the real (imaginary) time evolution of parametrized trial states via forward Euler time-stepping of the Wick rotated Schrödinger equation, thereby solving the Black-Scholes equation, and by extension, the heat equation \cite{fontanela2021quantum,miyamoto2021pricing}. Besides issues of ansatz selection and quantum complexity, time-stepping based on an explicit Euler method may be unstable, a limiting condition exacerbated by noise.

This paper is organized as follows. In \Cref{sec:sec2}, we outline general implicit time-stepping schemes for solving evolution partial differential equations and propose the use of a variational quantum solver to resolve the Laplacian operator iteratively. In \Cref{sec:sec3}, we apply the variational quantum algorithm to solving a heat or diffusion equation without source terms as a proof of concept. With that, we explore potential applications to more general evolution problems with non-linear source terms, including the reaction-diffusion (\Cref{sec:sec4}) and the Navier-Stokes equations (\Cref{sec:sec5}), where variables can be coupled through semi-implicit schemes.

\section{Theory} \label{sec:sec2}

Consider the second-order homogeneous evolution equation defined on the set $\Omega \times J$, where 
$\Omega \subset \mathbb R^d$ denotes a $d$-dimensional bounded spatial domain and $J = [0,T]$, where $T>0$ denotes a bounded temporal domain, as
\begin{alignat}{2}
\frac{\partial u(\vec x,t)}{\partial t} &= D \nabla^{2} u(\vec x,t) + f(\vec x,t), \qquad 
&& \mbox{in } \Omega \times J \label{eq:pde} \\
u(\vec x,0)&=u_{0}(\vec x),  && \mbox{in } \Omega \times\{t=0\},
\label{eq:initial}
\end{alignat}
where $u(\vec x,t)$ is a function of spatial vector $\vec x$ and time $t$, $D>0$ is the diffusion coefficient and $f$ is an unspecified source term. 
For now, Dirichlet and Neumann boundary conditions are applicable on the boundary $\Gamma := \partial\Omega = \Gamma_D\cup\Gamma_N$, respectively,
\begin{alignat}{2}
u&=g, \quad \mbox{in } \Gamma_{D} &&\times J, \label{eq:dirichlet} \\
\frac{\partial u}{\partial n}&=0, \quad \mbox{in }\Gamma_{N} &&\times J,
\label{eq:bc}
\end{alignat}
where $\partial/\partial n$ is the outward normal derivative on boundary $\Gamma$.

For a two-dimensional rectangular domain $\Omega = (x_L,x_R )\times (y_L,y_R) \subset \mathbb R^2$, partitioning the space-time domain $\Omega \times J$ yields the space-time grid points
\begin{align}
    \left(x_{i j}, t^{k}\right):=\left(x_{i}, y_{j}, t^{k}\right), \quad i=0,1, \ldots, n_{x}; j=0, 1, \ldots, n_{y};  k=0, 1, \ldots, n_{t},
\label{eqT3}
\end{align}
where $n_x$, $n_y$ and $n_t$ are prescribed positive integers, such that $x_i=x_L+i\cdot \Delta x$, $y_j=y_L+j \cdot \Delta y$, $t^k=k\cdot \Delta t$, $\Delta x=L_x/n_x$ , $\Delta y=L_y/n_y$  and $\Delta t=T/n_t$ , where $L_x=x_R-x_L$ and $L_y=y_R-y_L$. The discrete domain grid is denoted by $\Omega_d=\{(x_i,y_j ): n_x \in \{0,1,\ldots,n_x\}, n_y \in \{0,1,\ldots,n_y\} \}$ and boundary grid by $\Gamma_d$. 

The finite difference (FD) approximation for the second-order spatial derivative (5-point) of the Laplacian operator taken at $t=t^k$ is
\begin{align}
    \mathcal{A}u_{i, j}=-\delta_{x}(u_{i-1, j}-2 u_{i, j}+u_{i+1, j})-\delta_{y}(u_{i, j-1}-2 u_{i, j}+u_{i, j+1}),
\label{eqT4}
\end{align}
where $\delta_x :=D \Delta t/\Delta x^{2}$ and $\delta_y:=D \Delta t/\Delta y^{2}$ are diffusion parameters.

Using first-order FD for temporal derivative $(u_{ij}^{k+1}-u_{ij}^k)/\Delta t$ weighted by $\vartheta \in [0,1]$, the evolution \cref{eq:pde} can be expressed in vector shorthand as
\begin{align}
    (\mathcal{I}+\vartheta \mathcal{A}) u^{k+1}=[\mathcal{I}-(1-\vartheta) \mathcal{A}] u^{k}+\Delta t f^{k+\vartheta},
\label{eqT5}
\end{align}
where $\mathcal{I}$ is the identity matrix of the same size, $u^{k}=\left[u_{i j}^{k}\right]_{0 \leq i \leq n_{x}, 0 \leq j \leq n_{y}}$ and  $f^{k}=\left[f_{i j}^{k}\right]_{0 \leq i \leq n_{x}, 0 \leq j \leq n_{y}}$.

Depending on the choice of parameter $\vartheta$, actual time-stepping may follow an explicit (forward Euler) method ($\vartheta=0$), an implicit (backward Euler) method ($\vartheta=1$), a semi-implicit (Crank-Nicolson) method ($\vartheta=1/2$) or a variable-$\vartheta$ method \cite{lee2020variable}. The explicit method ($\vartheta=0$) is efficient for each time-step but is only stable if it satisfies the stability condition $v\leq 1/2$. The implicit (backward Euler) method ($\vartheta=1$) is unconditionally stable and first-order accurate in time ($\varepsilon \sim \Delta t$), which reads
\begin{align}
    (\mathcal{I}+\mathcal{A}) u^{k+1}=u^{k}+\Delta t f^{k+1}.
\label{eq:IE}
\end{align}
The semi-implicit Crank-Nicolson (CN) method ($\vartheta=1/2$) is popular as it is not only stable, but also second-order accurate in both space and time ($\varepsilon \sim \Delta t^2 $), which reads
\begin{align}
    \left(\mathcal{I}+\frac{\mathcal{A}}{2}\right) u^{k+1}=\left(\mathcal{I}-\frac{\mathcal{A}}{2}\right) u^{k}+\Delta t f^{k+1 / 2}
\label{eq:CN}
\end{align}
where $f^{k+1/2}=(f^{k+1}+f^k )/2$. However, the CN method may introduce spurious oscillations to the numerical solution for non-smooth data unless the algorithm parameters satisfy the maximum principle \cite{crank1947practical}.

\subsection{Variational Quantum Solver}
Here, we explore a variational quantum approach towards the solution of the evolution equation \eqref{eq:pde}. In addition to potential quantum speedup, a variational quantum algorithm could also benefit from data compression, where a matrix of dimension $N$ can be expressed by a quantum system with only $\log_2 N$ qubits, where $N$ is the size of the problem. Consider the Poisson equation, which is a time-independent form of Eq.~\eqref{eq:pde}, expressed as
\begin{align}
    -\nabla^2 u=f,\qquad \mbox{in } \Omega \subset \mathbb R.
    \label{eq:poissonEq}
\end{align}

The Laplacian operator $\nabla^2$ in one dimension can be discretized using the finite difference method in the $x$ direction into an $N$×$N$ coefficient matrix $A_{x,\beta}$ as
\begin{align}
    A_{x,\beta} = 
    \begin{bmatrix}
        1+\alpha_\beta & -1 & 0 & & & \cdots &  & 0 \\
        -1 & 2 & -1 & 0 & & \cdots  & & 0 \\
        0 & -1 & 2 & -1 & & \cdots  & & 0 \\
        \vdots &  &  & \ddots  & & & & \vdots \\
        0 & & \cdots & & 0 & -1 & 2 & -1  \\
        0 & & \cdots & & & 0 & -1 & 1+\alpha_\beta
    \end{bmatrix}.
    \label{eq:MxN}
\end{align}
where $\beta \in \{D,N\}$ refers to either the Dirichlet ($D$) or Neumann ($N$) boundary condition, and $\alpha_D=1$ and $\alpha_N=0$. This extends naturally to higher dimensions, for instance $A_{y,\beta}$ in the $y$ direction. 

A variational quantum solution is to prepare a state $\ket u$ such that $A \ket u$ is proportional to a state $\ket b$ in a way that satisfies Eq.~\eqref{eq:poissonEq}. To do that, a canonical approach \cite{Xu2019variational, huang2019near, bravoprieto2020variational} is to first decompose the matrix $A$ over the Pauli basis 
$\mathcal P_n = \{P_1\otimes \cdots \otimes P_n: \forall  i,P_i \in \{I,X,Y,Z\}  \}$ (where $X=\kbra 10 +\kbra 01$, $Y = i\kbra 10 -i \kbra 01$, and $Z=\kbra 00 -\kbra 11$ are the Pauli matrices and $I=\kbra 00 +\kbra 11$ is the identity matrix) as
\begin{align}
    A=\sum_{P \in \mathcal{P}_{n}} c_{P} P,
    \label{eq:pauli_decomposition}
\end{align}
where $c_P=\tr(PA)/2^n$ are the coefficients of $A$ in the Pauli basis. Using simple operators $\sigma_+ = \kbra 01$, $\sigma_- = \kbra 10$, the number of terms in the decomposition can be reduced to $2 \log N+1$ \cite{liu2021variational}. A more efficient approach, however, is to express $A$ as a linear combination of unitary transformations of simple Hamiltonians \cite{sato2021variational}. Accordingly, the decomposition of $A$ in one dimension can be written as \cite{ewe2021variational}
\begin{align}
    A_{x, \beta}=I^{\otimes n-1} \otimes(I-X)+S^{\dagger}\left[I^{\otimes n-1} \otimes(I-X)+I_{0}^{\otimes n-1} \otimes\left(X-a_{\beta} I\right)\right] S,
    \label{eq:ewe_decomposition}
\end{align}
where $I_0 = \kbra 00$  and $\beta \in \{D,N\}$, as before except here, $a_D=0$ and $a_N=1$. Here, $S$ is the $n$-qubit cyclic shift operator defined as
\begin{align}
S = \sum_{i=0}^{2^n-1}
\kbra{(i+1)\ \operatorname{mod}\ 2^n}i.
\label{eq:shiftP}
\end{align}

The expectation values of a Hamiltonian $H$ including the shift operator $S$ are evaluated by applying the unitary shift operator to the quantum state \cite{sato2021variational},
\begin{align}
    \bra \phi S^\dag H S \ket \phi = \bra{\phi'} H \ket{\phi'},
    \label{eq:shift_expectation}
\end{align}
where $\ket \phi$ is an arbitrary $n$-qubit state and $\ket {\phi'}= S \ket \phi$.
Note that Eq.~\eqref{eq:ewe_decomposition} can be re-written as
\begin{align}
    A_{x, \beta}=2 I^{\otimes n}-\underbrace{I^{\otimes n-1} \otimes X}_{H_{1}}+S^{\dagger}\big[-\underbrace{I^{\otimes n-1} \otimes X}_{H_{2}}+\underbrace{I_{0}^{\otimes n-1} \otimes X}_{H_{3}}-\underbrace{b_{\beta} I_{0}^{\otimes n-1} \otimes I}_{H_{4}}\big] S.
    \label{eq:A_decomposition}
\end{align}
Since expectation values of the identity operator are equal to 1, i.e.~$\bra \phi I^{\otimes n} \ket \phi = \bra{\phi'} I^{\otimes n} \ket{\phi'} = 1$, evaluating the expectation value of the operator $A_{x, \beta}$ requires only the evaluation of expectation values of the simple Hamiltonians $H_{1-4}$ ($H_{1-3}$ for Dirichlet boundary condition). The required number of quantum circuits is therefore limited to a constant $O(n^0)$ \cite{sato2021variational}. Similar decomposition expressions apply to problems of higher dimensions, including $A_{y,\beta}$ in the $y$ direction \cite{ewe2021variational}.

\begin{figure}
\centering
\begin{align*}
\Qcircuit @C=0.3em @R=.4em {
 &  & & \mbox{layer 1} 
& &  & & & & & & & &  &  & & & \mbox{layer $n$}  
\\~\\
& \lstick{\ket{0}}\qw & \gate{R_y(\theta^1_1)} & \ctrl{1} & \qw  & \qw & \qw & \qw & \qw & \qw &  & &  \ldots & & & &  \gate{R_y(\theta^n_1)} & \ctrl{1} & \qw & \qw   & \qw & \qw & \qw & \qw
\\
& \lstick{\ket{0}}\qw & \gate{R_y(\theta^1_2)} & \targ    & \ctrl{1} & \qw   & \qw & \qw & \qw & \qw &  & &  \ldots & &  & & \gate{R_y(\theta^n_2)} & \targ    & \ctrl{1} & \qw   & \qw & \qw & \qw & \qw
\\
& \lstick{\ket{0}}\qw & \gate{R_y(\theta^1_3)} & \qw & \targ & \ctrl{1} & \qw & \qw & \qw & \qw & & & \ldots & & & &   \gate{R_y(\theta^n_3)} & \qw      & \targ    &   \ctrl{1} & \qw & \qw & \qw & \qw 
\\
& &  & & & \targ & \qw & & & & & & & & &  & & & & \targ & \qw & 
\\~\\~\\
& \lstick{\vdots} & \vdots & & &  &  \ddots  & & & & & & & & & & \vdots  & & & & \ddots &
\\~\\~\\~\\
& & & & & & &  \ctrl{1} & \qw  & & & & & & & &  & & & & & \ctrl{1} & \qw &
\\
& \lstick{\ket{0}}\qw & \gate{R_y(\theta^1_n)} & \qw      & \qw    &  \qw & \qw & \targ & \qw & \qw &  &  & \ldots & & & &   \gate{R_y(\theta^n_n)} & \qw      & \qw    &  \qw & \qw & \targ & \qw & \qw 
\gategroup{3}{3}{14}{8}{.3em}{.}
\gategroup{3}{17}{14}{22}{.3em}{.}\\
}
\end{align*}
\caption{Schematic of the hardware-efficient ansatz used in this study.}
\label{fig:hardware_efficient_ansatz}
\end{figure}
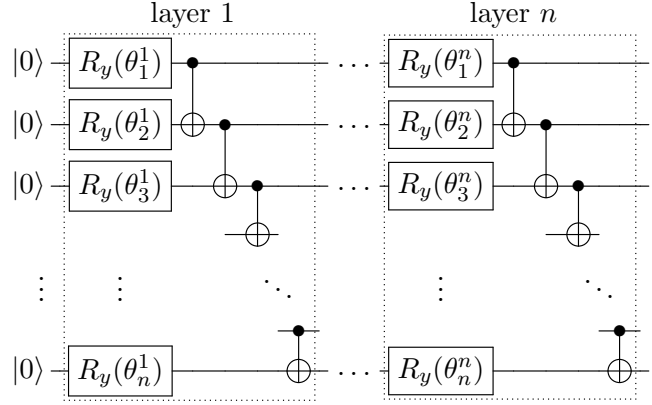

Once the matrix $A$ is decomposed, a parameterized quantum state $\ket{\psi(\theta)}$ is prepared using an ansatz represented by a sequence of quantum gates $U(\theta)$ parameterized by $\theta$ applied to a basis state $\ket 0^{\otimes n}$, such that $\ket {\psi(\theta)} = U(\theta) \ket 0^{\otimes n}$. Here, we use a hardware-efficient ansatz consisting of multiple layers of $R_Y$ gates across $n$ qubits entangled by controlled-$X$ gates (see \Cref{fig:hardware_efficient_ansatz}). For the source term $f$ in \eqref{eq:poissonEq}, a quantum state $\ket b$ is prepared by encoding a real vector with the unitary $U_b$, such that $\ket b = U_b \ket 0^{\otimes n}$. Depending on the actual input, conventional amplitude encoding methods \cite{mottonen2004transformation,shende2006synthesis} may introduce a global phase that must be corrected by its complex argument for computing in the real space.

With $\ket{\psi(\theta)}$ and $\ket b$, the cost function $E$ can be optimized in terms of $A$ as \cite{sato2021variational}
\begin{align}
    E(r(\theta), \theta)=-\frac{1}{2} \frac{\left|\bket{\psi(\theta)}{b}\right|^{2}}{\langle\psi(\theta)|A| \psi(\theta)\rangle},
    \label{Eq:costE}
\end{align}
where 
$\ket{\psi(\theta),b}:=
(\ket 0 \ket{\psi(\theta)}  + \ket 1 \ket b)/\sqrt{2}
$. The norm of the state vector $\ket {\psi(\theta)}$ is represented by $r\in \mathbb R$, where
\begin{alignat}{2}
     r(\theta)&:=\frac{\left|\bket{\psi(\theta)}{b}\right|}{\bra{\psi(\theta)}A\ket{\psi(\theta)}} \nonumber \\
     &=\frac{\left|\bra{\psi(\theta),b}X \otimes I^{\otimes n}\ket{\psi(\theta),b}-i\bra{\psi(\theta),ib} X \otimes I^{\otimes n}\ket{\psi(\theta),ib}\right|}{\bra{\psi(\theta)}A\ket{\psi(\theta)}}.
    \label{Eq:ropt}
\end{alignat}

Using classical optimization tools, the cost function \eqref{Eq:costE} is minimized with $\theta$ updated iteratively until convergence is reached. The optimization process follows either a gradient-based or gradient-free approach, depending on how the gradient of the cost function is evaluated. A gradient-free optimizer is guided by an estimate of the inverse Hessian matrix, whereas a gradient-based optimizer by the partial derivative of the cost function $E$ with respect to parameters $\theta$, i.e.~$\partial E /\partial \theta$, which can be evaluated by a quantum computer (for details, see \cite{sato2021variational,ewe2021variational}). Regardless of the choice of gradient optimizer used, the optimization routine halts when the cost error falls under a convergence threshold ($\epsilon < \epsilon_{\tol}$) whence the parameters are at optimum $\theta=\theta_\opt$. The converged solution vector $\ket u = r_{\opt} \ket{\psi(\theta_{\opt}}$ satisfies \cite{bravoprieto2020variational}
\begin{align}
    r_{\opt}A\left|\psi\left(\theta_{\opt}\right)\right\rangle=|b\rangle
    \label{eq:r_opt_psi_theta_opt}
\end{align}
where $r_{\opt}=r(\theta_{\opt} )$ is the norm of the solution to the Poisson equation \eqref{eq:poissonEq}.

In this study, we propose to solve the evolution equation \eqref{eq:pde} through successive time-stepping of the quasi-steady Poisson equation using a variational quantum algorithm. Using a parameter set $\theta^k$ obtained at time-step $k$, we encode a normalized source state $\ket{\hat{b}^k}:=
\ket{b}/\sqrt{\bket{b}{b}}
$ from $\ket{b(\theta^k)}$ and seek an implicit solution to
\begin{align}
    r^{k+1}A\left|\psi\left(\theta^{k+1}\right)\right\rangle=\left|\hat{b}^{k}\right\rangle,
    \label{eq:rk+1psi}
\end{align}
where $\theta^{k+1}=\theta_\opt (t^{k+1} )$ is the parameter set and $r^{k+1}=r_\opt (\theta^{k+1})$ is the norm at next time-step $k+1$. This process is then iterated in time up to $n_t$ number of time-steps as desired (see Algorithm \ref{algo01}).

\begin{minipage}{0.95\linewidth}
\noindent
\begin{algorithm}[H]
	\caption{Variational Quantum Evolution Equation Solver}
	\label{algo01}
	\begin{algorithmic}[1]
	    \State {Initialize} $\ket{\psi(\theta)}$
		\For {$k \in [0,n_t)$}
			\State {Encode} $\ket{\hat{b}^{k}} \gets \ket{u^{k}}$		 
			\While {$\varepsilon > \varepsilon_{\tol}$}
				\State {Evaluate} $E(\theta ^k)$  (Eq.~\ref{Eq:costE}) (and optionally $(\partial E/\partial \theta)$) on a quantum computer 
				\State Update $\theta ^k$ using classical optimization 
			\EndWhile
			\State {Update} $\ket{u^ {k+1}} \gets r(\theta^k)\ket{\psi(\theta^k)}$ (Eq.~\ref{Eq:ropt})
		\EndFor
		\State {Return} $\ket{u}$
	\end{algorithmic} 
\end{algorithm}
\end{minipage}
\vspace{\baselineskip}\linebreak

In this study, the variational quantum algorithm is implemented in Pennylane (Xanadu) \cite{bergholm2018pennylane} using a statevector simulator with the Qulacs \cite{Suzuki2021qulacs} plugin as a backend for quantum simulations, and the L-BFGS-B optimizer for parametric updates. Amplitude encoding is carried out via the standard Mortonnen state preparation template \cite{Schuld2019} with custom global phase correction. For hardware emulation via the QASM simulator (Qiskit), we refer the reader to the excellent cost-sampling analysis of Sato et al.~\cite{sato2021variational}.

\section{Applications to the Heat/Diffusion Equation}  \label{sec:sec3}

Consider the following one-dimensional heat or diffusion equation without a source term
\begin{alignat}{2}
\frac{\partial u}{\partial t} & =D \frac{\partial^{2} u}{\partial x^{2}}, 
\qquad && \mbox{in }\Omega \times J \\
u & =u_{0}(x), && \mbox{in }\Omega \times\{t=0\}.
\label{eq:one_dim_heat_diff}
\end{alignat}

Dirichlet conditions are applied on the boundaries of a 1D domain $\Omega=(x_L,x_R)\subset \mathbb R$, where $u(x_L,t)=g_L (t)$ and $u(x_R,t)=g_R (t)$, such that the boundary vector $u_D=(g_L, 0,\ldots, 0,g_R )$ is known for all $t$.

To solve Eq.~\eqref{eq:one_dim_heat_diff}, the variational quantum evolution algorithm (Algorithm \ref{algo01}) can be employed with a suitable time-stepping scheme \eqref{eqT5}. For the implicit Euler (IE) method \eqref{eq:IE}, the matrix $A$ and source state $\ket{b(\theta^k)}$ can be decomposed into
\begin{align}
    \begin{gathered}
A = I^{\otimes n}+\delta_{x}A_{x, D}, \\
\ket{b^k} = \ket{u^k} + \delta_{x}\ket{u_{D}^{k+1}},
\end{gathered}
\label{eq:decomposed_into}
\end{align}
where $\ket{u^k} = r^k\ket{\psi\left(\theta^{k}\right)}$.

For the Crank-Nicolson (CN) method \eqref{eq:CN}, it follows that the $\mathcal Au^k$ term, which carries a small but non-trivial evaluation cost, can be eliminated using the source state of the previous time-step $k-1$, leading to
\begin{align}
    \begin{gathered}
A = 2I^{\otimes n} + \delta_{x}A_{x, D}, \\
\ket{b^k} = 4\ket{u^k} - 2\delta_{x}\left(\ket{b^{k-1}} + \ket{\bar{u}_{D}^{k+1/2}}\right).
\end{gathered}
\label{eq:leading_to}
\end{align}
where $2\left|\bar{u}_{D}^{k+1/2}\right\rangle:=\left(\left|u_{D}^{k+1}\right\rangle+\left|u_{D}^{k}\right\rangle\right)$. Here, the presence of a $k-1$ term in $\left|b^{k-1}\right\rangle$ is not unexpected due to temporal finite differencing at second-order accuracy. 

For a space-time domain $\Omega\times J\in[0,1] \times [0,1]$, let the number of time-steps be $n_t=20$ and the spatial intervals be $n_x=2^n+1$, where $n$ is the number of qubits, and $\delta_x=1$ is the diffusion parameter. We employ the Dirichlet boundary condition with boundary values $(g_L,g_R )=(1,0)$ and initial values $u_0= \vec 0$. With initial random parameters $\theta^0 \in [0,2 \pi]$, we run a limited-memory Broyden-Fletcher-Goldfarb-Shanno boxed (L-BFGS-B) optimizer \cite{shanno1970conditioning,goldfarb1970family,fletcher1970new,broyden1970convergence} to optimize $\theta$ with absolute and gradient tolerances set at $10^{-8}$.

\begin{figure}
\centering
  \includegraphics[width=0.9\linewidth]{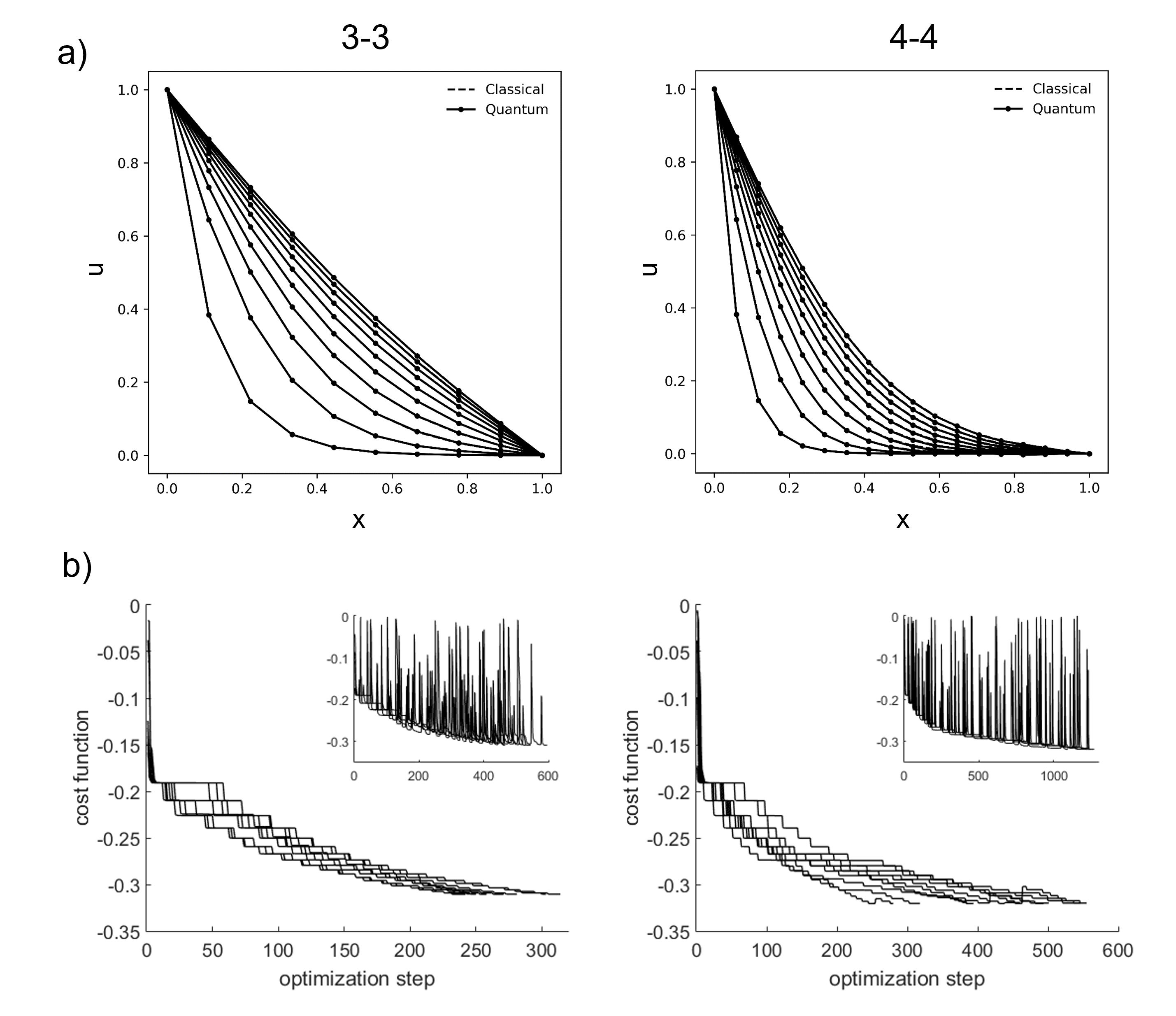}
\caption{(a) Implicit variational quantum solutions to a 1D heat conduction or diffusion problem in time-increments of 0.1, with boundary values $\{g_L,g_R \}=\{1,0\}$, initial values $u_0=\bar{0}$ and diffusion parameter $\delta_x=1$. (Left) Qubit–layer count $n\mh l=3\mh 3$ and time-averaged trace error $\bar{\varepsilon}_{\tr}=0.0008$; (Right) $n\mh l=4\mh 4$, $\bar{\varepsilon}_{\tr}=0.0025$. (b) Cost function vs.~number of optimization steps for 10 runs. Inset: Input parameters are re-initialized randomly, $\theta \in [0,2 \pi ]$, before each time-step for 5 runs.}
\label{fig:fig1}
\end{figure}

\Cref{fig:fig1}a compares solutions obtained from the variational quantum solver \eqref{eq:decomposed_into} and classical methods to a 1D heat or diffusion problem in time-increments of $0.1$, where the number of qubits and ansatz layers expressed as a set $n\mh l$, are $3\mh 3$ and $4\mh 4$. Here we define the time-averaged trace error $\bar \epsilon_{\mathrm{tr}}$ as 
\begin{align}
    \bar{\varepsilon}_{\mathrm{t r}}:=\frac{1}{n_{t}} \sum_{k=0}^{n_{t}-1} \sqrt{\left.1-\left|\langle \psi\left(\theta^{k}\right)\right| \hat{u}^{k}\right\rangle\left.\right|^{2}},
    \label{eq:time_averaged_trace_error}
\end{align}
where $\left|\hat{u}^{k}\right\rangle:=\left|u^{k}\right\rangle / \sqrt{\left\langle u^{k} \mid u^{k}\right\rangle}$ is the normalized classical solution vector at time $k$. The trace errors of solutions shown in \Cref{fig:fig1}a are 0.0008 and 0.0025 for $n\mh l$ sets of $3\mh 3$ and $4\mh 4$ respectively.

\Cref{fig:fig1}b shows how the cost function $E$ depends on the number of optimization steps for $n\mh l$ of $3\mh 3$ and $4\mh 4$ ($10$ sampled runs each). Each distinct step in $E$ represents sequential optimization from solution $\ket {\psi(\theta^k)}$ at time-step $k$ towards the solution $\ket{\psi(\theta^{k+1})}$ at $k+1$. For small time-step $\Delta t$, $\theta ^k$ provides a good initial parameter set for solving optimization step $k+1$. If the ansatz parameters were re-initialized randomly $\theta^k\in[0,2\pi]$ before each time-step, significantly more optimization steps would be required on average for convergence for each run (see Fig.~\ref{fig:fig1}b, inset). 

\subsection{Time complexity}
Here we briefly examine the time complexity of the quantum algorithm excluding the classical computing components. Following the analysis of the variational Poisson solver \cite{sato2021variational}, the time complexity of the proposed variational evolution equation solver per time-step reads
\begin{align}
    T \sim \mathcal{O}\left(\bar{T}_{\eval}\left(\frac{l+e+n^{2}}{\varepsilon^{2}}\right)\right),
    \label{eq:time_complexity}
\end{align}
where the terms within the inner parentheses indicate the time complexity of the state preparation scaling as $\mathcal{O}(l+e+n^2 )$, which consists of the ansatz depth $l$, the encoding depth $\mathcal{O}(n^2)$ \cite{Araujo2021}, and the depth $O(n^2)$ of the circuit needed to implement the $n$-qubit cyclic shift operator, and that of the number of shots $\mathcal{O}(\varepsilon^{-2})$ required for estimation of expectation values up to a mean squared error of $\varepsilon^2$. The required number of quantum circuits depends on the boundary conditions applied (3 for periodic, 4 for Dirichlet and 5 for Neumann conditions), scaling only as $\mathcal{O}(n^0 )$. $
\bar T_{\eval}$ is the time-averaged number of function evaluations,
\begin{align}
    \bar{T}_{\eval}:=\frac{1}{n_{t}} \sum_{k=0}^{n_{t}-1} T_{\eval},
    \label{eq:time_averaged_number_of_function_evaluations}
\end{align}
where $T_{\eval}$ is the sum of function evaluations required for a run with $n_t$ time-steps. Using a gradient-based optimizer, the time complexity for gradient estimation via quantum computing would scale as the ansatz volume $\mathcal{O}(nl)$ representing the number of quantum circuits required for parameter shifting. Otherwise, with a gradient-free optimizer, the time complexity simply contributes towards $\bar T_{\eval}$ as additional function evaluations required to evaluate the Hessian for gradient descent.

\begin{figure}
\centering
  \includegraphics[width=0.9\linewidth]
  {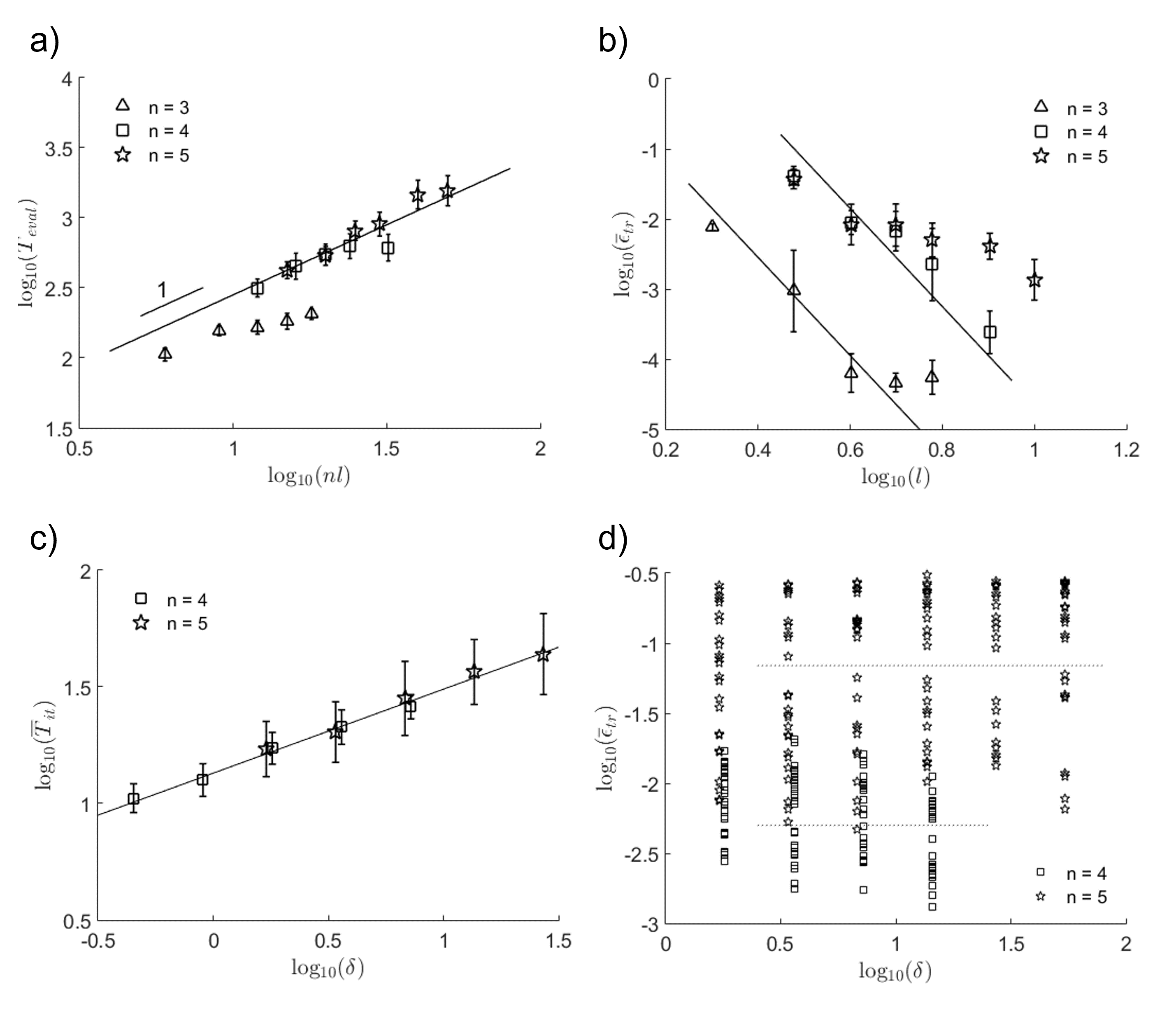}
\caption{Logarithmic plots of time-averaged (a) number of function evaluations $\bar{T}_{\eval}$ vs.~number of parameters $nl$, (b) trace error $\bar{\varepsilon}_{\tr}$ vs.~number of layers $l$, (c) number of iterations $\bar{T}_{\it}$ and (d) trace error $\bar{\varepsilon}_{\tr}$ vs.~diffusion parameter $\delta$ for $l=n$ up to $T=1$. Each data point and error bar represents, respectively, the mean and standard deviation out of 25 runs.}
\label{fig:fig2}
\end{figure}

To see if the time complexity for gradient-free optimization scales as $\mathcal{O}(nl)$, we plot the time-averaged number of function evaluations $\bar T_{\eval}$ against the number of parameters $nl$ (Fig.~\ref{fig:fig2}a). Indeed, we found that $\bar T_{\eval}$  scales reasonably with $nl$ (see trendline of slope 1), despite apparent tapering at higher $l$. Fig.~\ref{fig:fig2}b shows that the time-averaged trace error $\bar{\varepsilon}_{\tr}$ decreases with circuit depth $l$, even for over-parameterized quantum circuits where the number of layers exceeds the minimum required for convergence, $l_{\mathrm{min}} := 2^n/n$ \cite{patil2022variational}. For low grid resolution $n = 3$, the  trace error is limited to a minimum of $\sim10^{-4}.$

For deep and wide quantum circuits, the increase in optimization time is exacerbated by the presence of barren plateaus, or vanishingly small gradients in the energy landscape, where re-initialization can leave one trapped at a position far removed from the minimum \cite{Huembeli2021,McClean2018,cerezo2021cost}. Conversely, short time-steps lead to efficient solution trajectories that remain close to the local cost minima, leading to significant reduction in optimization times. To verify this, we conduct numerical simulations varying the diffusion parameter $\delta$ with $l=n$ up to time $T=1$. \Cref{fig:fig2}c shows that the number of iterations, or required optimization steps, per time-step increases linearly with $\delta$. Close inspection of the time-averaged trace distance shows bimodal distributions at higher $\delta$, which separates success and failure during convergence towards the global minimum (see Fig.~\ref{fig:fig2}d, dotted lines), resembling local minima traps due to poor optimization or expressivity of ans\"{a}tze \cite{ewe2021variational,Wierichs2020}.

\subsection{Discretization error}
\begin{figure}
\centering
  \includegraphics[width=0.9\linewidth]
  {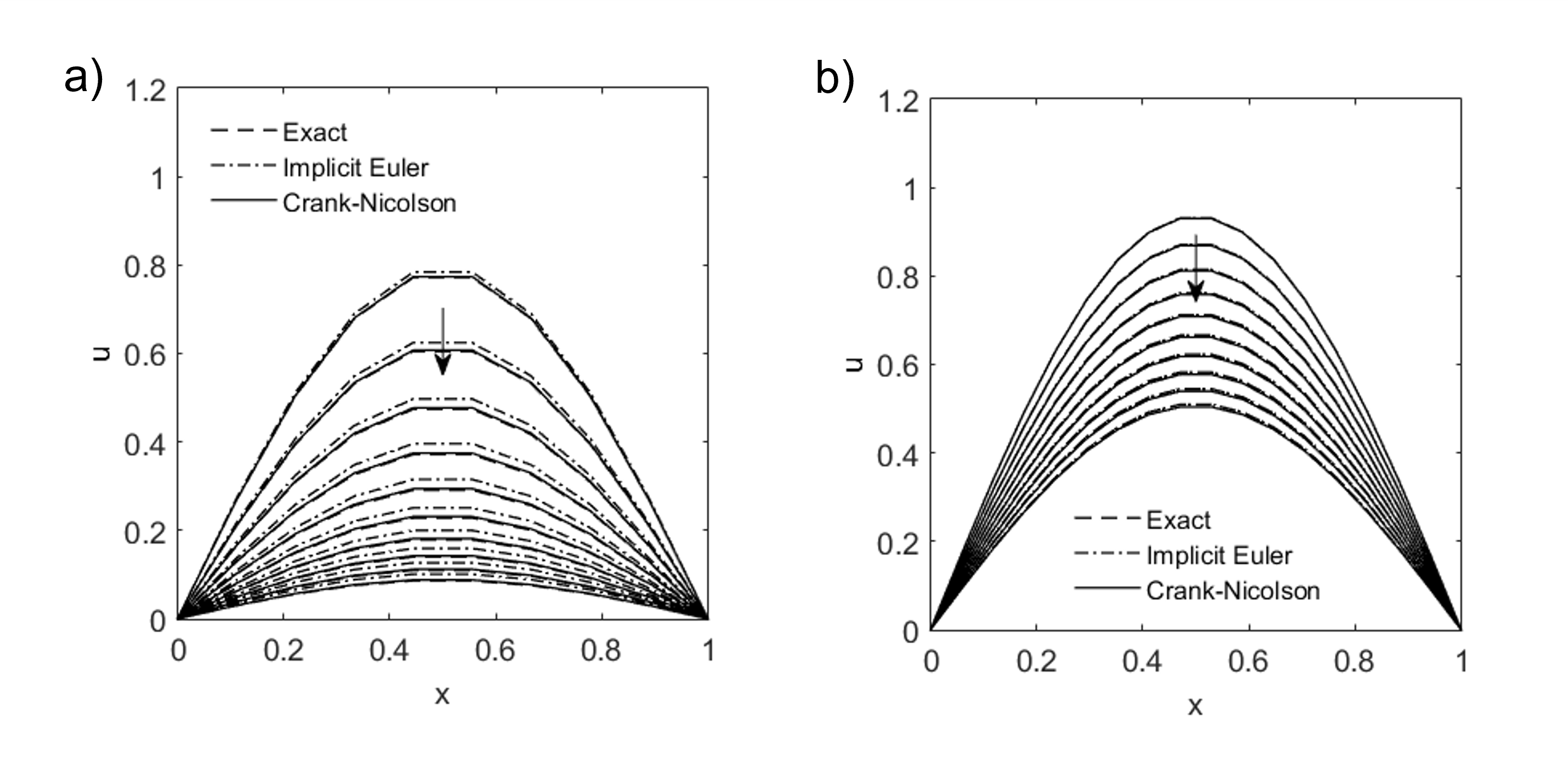}
\caption{Variational quantum solutions to a 1D heat conduction or diffusion problem for qubit-layers (a) $n\mh l=3\mh 3$ and (b) $n\mh l=4\mh 4$ in time-increments of 0.1 using implicit Euler and Crank-Nicolson schemes, with Dirichlet boundary values $(g_L,g_R )=(0,0)$, initial values $u_0=\sin{(\pi x)}$ and diffusion parameter $\delta_x=1$. Dashed lines denote exact solutions.}
\label{fig:fig3}
\end{figure}

Time evolution can be at a higher order, specifically for the Crank-Nicolson method. The problem statement is identical to the previous one, except with Dirichlet boundary values
$(g_L, g_R) = (0,0)$
and the initial condition $u_0 = \sin(\pi x/L_x$), where we use $L_x=1$ as the spatial length of the domain. This admits an exact analytical solution,
\begin{align}
    u(x, t)=\sin \left(\frac{\pi x}{L_{x}}\right) \exp \left[-D t\left(\frac{\pi}{L_{x}}\right)^{2}\right].
    \label{eq:exact_analytical_solution}
\end{align}
\Cref{fig:fig3} compares variational quantum and exact solutions using implicit Euler and Crank-Nicolson (CN) schemes. The discretization error for the higher-order CN scheme is reduced significantly, especially at lower grid resolution ($n=3$). Although the complexity costs for both methods \eqref{eq:decomposed_into} and \eqref{eq:leading_to} are similar, note however that the CN method may introduce spurious oscillations for non-smooth data \cite{lee2020variable}, an issue which may be exacerbated by quantum noise \cite{cincio2021machine}. 

\subsection{Higher dimensions}

The preceding analysis can be extended to higher dimensions. Consider the following two-dimensional heat or diffusion equation in $\Omega\times J$, where $\Omega =(x_L,x_R )\times (y_L,y_R) \subset \mathbb R^2$:
\begin{alignat}{2}
\frac{\partial u}{\partial t} & =D \left(\frac{\partial^{2} u}{\partial x^{2}} + \frac{\partial^{2} u}{\partial y^{2}}\right), \qquad && \mbox{in }\Omega \times J, \\
u & =u_{0}, && \mbox{in }\Omega \times\{t=0\}.
\label{eq:two_dim_heat_diff}
\end{alignat}

Under the implicit Euler scheme \eqref{eq:IE}, the matrix $A$ and source state $\ket{b^k}$ can be decomposed into

\begin{align}
    \begin{gathered}
A = I^{\otimes n} + \delta_{x}A_{x, D} + \delta_{y}A_{y, D}, \\
\left|b^k\right\rangle=\left|u^k\right\rangle+\delta_{x}\left|u_{x,D}^{k+1}\right\rangle+\delta_{y}\left|u_{y,D}^{k+1}\right\rangle.
\end{gathered}
\label{eq:decomposed_2d}
\end{align}

Dirichlet conditions are applied on the boundaries, where $u(x_{L,R},y,t) = g_{x_{L,R}}(y,t)$ and $u(x,y_{L,R},t) = g_{y_{L,R}}(x,t)$. Let the number of spatial grid intervals be $n_x = 2^{m_x} + 1$ and $n_y = 2^{m_y} + 1$, where $m_x$ is the number of qubits allocated to the $x$ grid, $m_y$ to the $y$ grid, and $n=m_x+m_y$ is the total number of qubits. Accordingly, $A$ is decomposed in terms of simple Hamiltonians in $x$ and $y$ as
\begin{multline}
    A_{x, \beta}=2 I^{\otimes n}-\underbrace{I^{\otimes n-1}\otimes X}_{H_{1}}+ \\
    S^{\dagger}_{[0,m_x)}\big[-\underbrace{I^{\otimes n-1}\otimes X}_{H_{2}} 
    +\underbrace{I^{\otimes m_y}I_{0}^{\otimes m_x-1}\otimes X}_{H_{3}}-\underbrace{b_{\beta}I^{\otimes m_y} I_{0}^{\otimes m_x-1} \otimes I}_{H_{4}}\big] S_{[0,m_x)}, 
    \label{eq:A2a_decomposition}
\end{multline}
\begin{multline}
    A_{y, \beta}=2 I^{\otimes n}-\underbrace{I^{\otimes m_y-1}\otimes X\otimes I^{\otimes m_x}}_{H_{1}}+ \\
    S^{\dagger}_{[m_x,n)}\big[-\underbrace{I^{\otimes m_y-1}\otimes X\otimes I^{\otimes m_x}}_{H_{2}}+\underbrace{I_{0}^{\otimes m_y-1}\otimes X\otimes I^{\otimes m_x}}_{H_{3}}-\underbrace{b_{\beta}I_{0}^{\otimes m_y-1} \otimes I^{\otimes m_x+1}}_{H_{4}}\big] S_{[m_x,n)},
    \label{eq:A2b_decomposition}
\end{multline}
where $H_{1-4}$ are simple Hamiltonians to be evaluated ($H_{1-3}$ for Dirichlet boundary condition). 

\begin{figure}
\centering
  \includegraphics[width=0.9\linewidth]
  {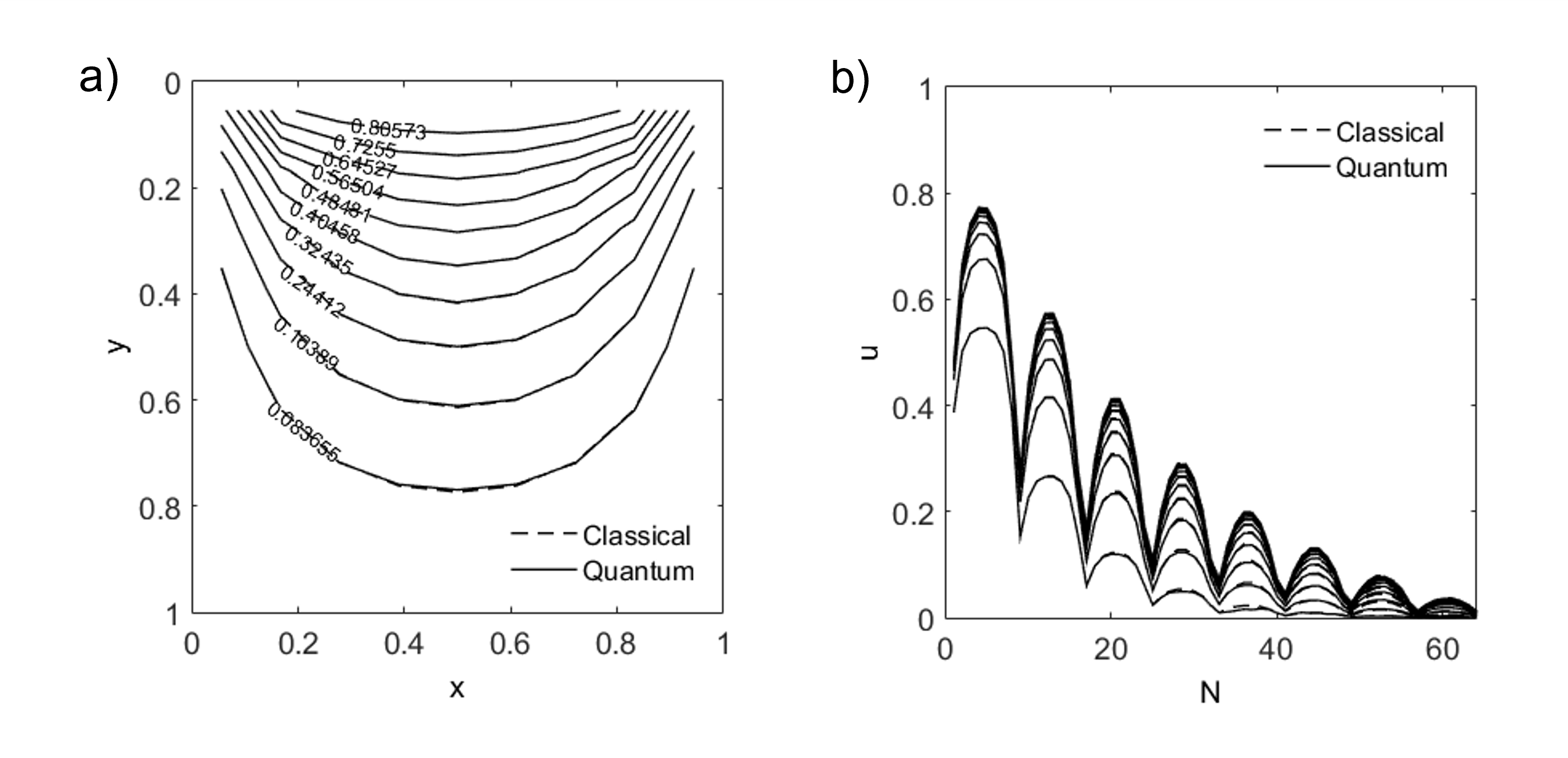}
\caption{(a) Contour solution plot of 2D heat conduction or diffusion problem at time $T = 1$ on a $8 \times 8$ x-y square grid (qubit-layer $m_x\mh m_y\mh l=3\mh 3\mh 6$) with Dirichlet boundary values $(g_{x_L},g_{x_R}) = (0,0)$ and $(g_{y_L},g_{y_R})=(1,0)$, initial values $u_0=\vec 0$, $n_t = 20$ and diffusion parameters $\delta_x=\delta_y=1$. (b) Solution vectors in time-increments of 0.1.}
\label{fig:fig4}
\end{figure}

\Cref{fig:fig4} shows solution snapshots to a 2D heat conduction or diffusion problem taken at time $T = 1$ with Dirichlet boundary values $(g_{x_L},g_{x_R}) = (0,0)$ and $(g_{y_L},g_{y_R})=(1,0)$, initial values $u_0=\vec 0$, $n_t = 20$ and diffusion parameters $\delta_x=\delta_y=1$. Results obtained from variational quantum solver agree with classical solutions with time-averaged trace errors of up to $10^{-2}$. 

\section{Applications to the Reaction-Diffusion Equations}  \label{sec:sec4}
Here, we extend applications of our variational quantum solver to evolution equations with non-trivial source terms. Consider a two-component homogeneous reaction-diffusion system of equations 
\begin{alignat}{2}
\frac{\partial \mathbf{u}(\vec x,t)}{\partial t} &= \mathbf{D} \nabla^{2} \mathbf{u}(\vec x,t) + \mathbf{f} (\vec x,t), \qquad 
&& \mbox{in } \Omega \times J, \label{eq:rd1} \\
\mathbf{u}(\vec x,0)&=\mathbf{u_0}(\vec x),  && \mbox{in } \Omega \times\{t=0\},
\label{eq:rd_initial}
\end{alignat}
where $\mathbf{u}=[u_1,u_2]^T$ is a concentration tensor, $\mathbf{D}=\operatorname{diag}[D_1,D_2]^T$ is a diffusion tensor and $ \mathbf{f} = [f_1(u_1,u_2), \allowbreak f_2(u_1,u_2)]^T$ is a coupled reaction source term. First proposed by Turing \cite{Turing1990}, the reaction-diffusion equations are useful for understanding pattern formation and self-organization in biological and chemical systems \cite{VanGorder2021,Kondo2010}, such as morphogenesis \cite{Gierer1972} and autocatalysis \cite{Gray1983}.

Here, we propose a \textit{semi-implicit} time-stepping scheme, whereby the coupled, non-linear source term is solved at the current time-step $k$. With explicit source term $f^k$, the implicit Euler scheme \eqref{eq:IE} reads
\begin{align}
    (I+\mathcal{A}) u^{k+1}=u^{k}+\Delta t f^{k}.
\label{eq:rd_IE}
\end{align}
The two-component tensor $\mathbf{A}=[A_1,A_2]^T$ and source state $\mathbf{b}=[b_1,b_2]^T$ can then be decomposed into
\begin{align}
    \begin{gathered}
\mathbf{A}=I^{\otimes n}+\mathbf{\delta_{x}A_{x, D}}, \\
\left|\mathbf{b^k}\right\rangle=\left|\mathbf{u^k}\right\rangle+\mathbf{\delta_{x}}\left|\mathbf{u_{D}^{k+1}}\right\rangle+\Delta t\left|\mathbf{f^k}\right\rangle,
\end{gathered}
\label{eq:decomposed_rd_IE}
\end{align}
where $\mathbf{\delta_x} = 2^{2n}\Delta t\mathbf{D}$ is the two-component diffusion parameter vector. With a linear Hermitian source matrix $f$, a fully implicit time-stepping scheme becomes available (Appendix \ref{appendix:implicit}).

\subsection{Implementation}
The semi-implicit variational quantum solver solves for the Laplacian for each component using a quantum computer and the solution vectors are explicitly coupled through source terms prior to re-encoding in preparation for the next time-step (see Algorithm \ref{algo02}).

\begin{minipage}{0.95\linewidth}
\noindent
\begin{algorithm}[H]
	\caption{Variational Quantum Solver for Reaction-Diffusion Equations (Semi-implicit)}
	\label{algo02}
	\begin{algorithmic}[2]
	    \State {Initialize} $\mathbf{\ket{\psi}(\theta_i)}$ for each component $i$
		\For {$k \in [0,n_t)$}
			\State {Encode} $\mathbf{\ket{\hat{b}^{k}}} \gets \mathbf{\ket{u^{k}}}$ (Eq.~\ref{eq:decomposed_rd_IE})		 
			\While {$\varepsilon_i > \varepsilon_{\tol}$}
				\State {Evaluate} $E_i(\theta_i^k)$ on a quantum computer 
				\State Update $\theta_i^k$ using classical optimization 
			\EndWhile
			\State {Update} $\mathbf{\ket{u^ {k+1}}} \gets \mathbf{r(\theta^k)} \mathbf{\ket{\psi(\theta^k)}}$
		\EndFor
		\State {Return} $\ket{\mathbf{u}}$
	\end{algorithmic} 
\end{algorithm}
\end{minipage}
\vspace{\baselineskip}\linebreak

\subsection{1D Gray-Scott Model}
The Gray-Scott model \cite{Gray1983} was originally conceived to model chemical reactions of the type $U + 2V \rightarrow 3V$, $V \rightarrow P$, where $U$, $V$ and $P$ are chemical species with reaction term
\begin{align}
    \mathbf{f(u)}=\begin{bmatrix}
        k_1(1-u_1)-u_1u_2^2 \\
        -(k_1+k_2)u_2+u_1u_2^2
    \end{bmatrix},
\label{eq:gs_f}
\end{align}
where $k_1$ and $k_2$ are kinetic rate constants.

An interesting class of Gray-Scott solutions involves periodic splitting of chemical wave pulses \cite{lee2020variable,Zegeling2004}. Here, we conduct a pulse splitting numerical experiment under limited spatial and temporal resolutions, using input parameters $\mathbf{D}=[10^{-4},10^{-6}]^T$, $k_1 = 0.04$ and $k_2 = 0.02$, with a \textit{mid-pulse} initial condition and Dirichlet boundary conditions as
\begin{alignat}{3}
u_1(x,0) = 1 - \frac{1}{2}\sin^{100}(\pi x), \qquad && u_2(x,0) = \frac{1}{4}\sin^{100}(\pi x), \qquad && x \in (0,1), \\
u_1(0,t) = u_1(1,t) = 1,  \qquad && u_2(0,t) = u_2(1,t) = 0, \qquad && t \in [0,T],
\label{eq:gs_param}
\end{alignat}
where time $t$ extends up to $T=600$ on $dt = 0.5$.

\begin{figure}
\centering
  \includegraphics[width=0.9\linewidth]
  {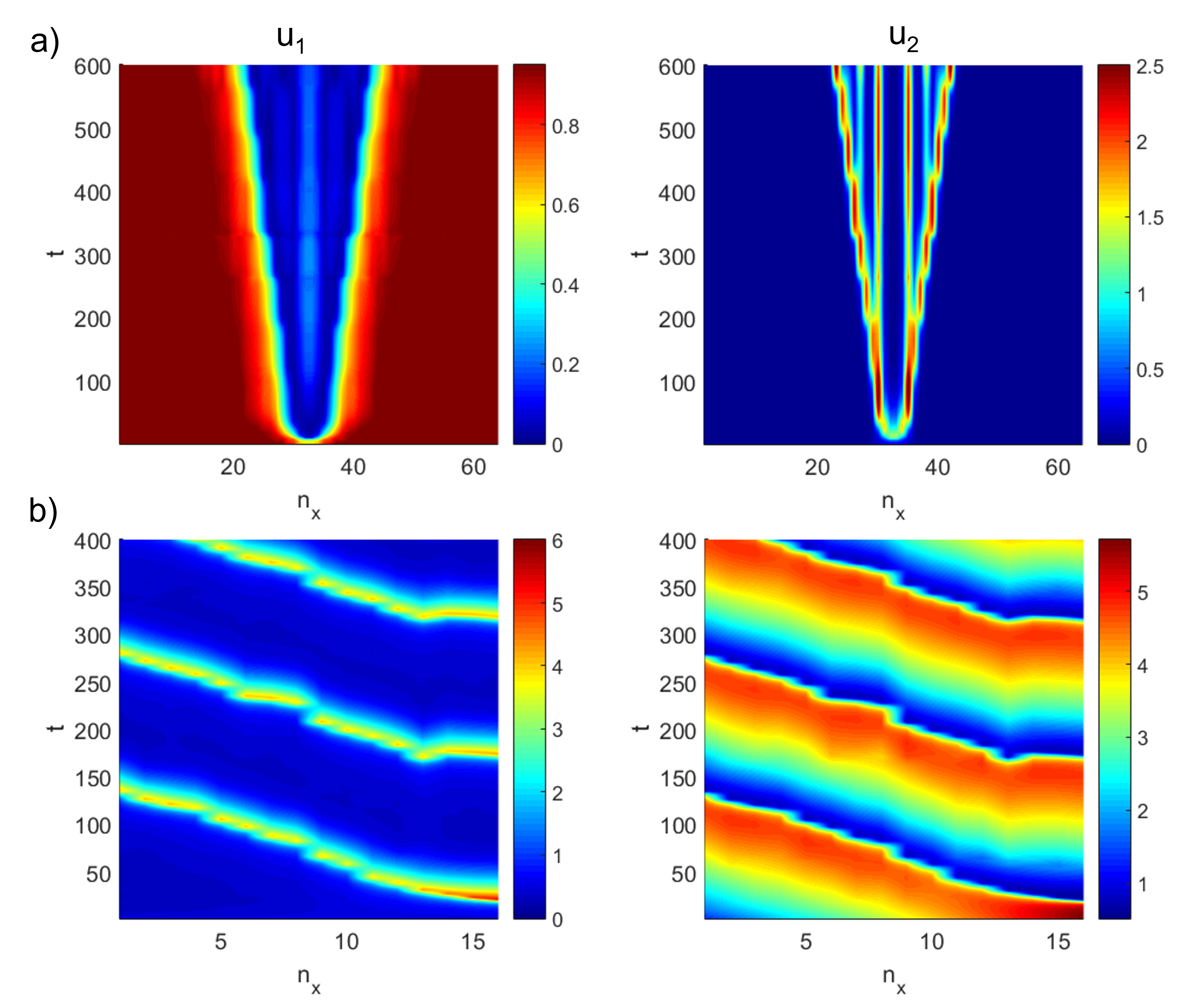}
\caption{Space-time solutions of (a) mid-pulse wave-splitting in 1D two-component Gray-Scott model obtained using semi-implicit variational quantum reaction-diffusion solver on $2^{6}=64$ grid points up to $T=600$, for chemical species $u_1$ (left) and $u_2$ (right). Parameters include $\mathbf{D}=[10^{-4},10^{-6}]^T$, $k_1 = 0.04$, $k_2 = 0.02$ and $dt=0.5$. (b) Traveling wave solutions for 1D Brusselator model on $2^{4}=16$ grid points up to $T=400$ on Neumann boundary conditions. Parameters include $\mathbf{D}=[10^{-4},10^{-4}]^T$, $k_1 = 3$, $k_2 = 1$ and $dt=0.5$.}
\label{fig:fig5}
\end{figure}

\Cref{fig:fig5}a shows how an initial mid-pulse can spontaneously and periodically split in space and time, a phenomenon captured using variational quantum diffusion reaction solver (see Algorithm \ref{algo02}) even on relatively low spatial resolutions. 

\subsection{1D Brusselator Model}
So far, we have been looking at only Dirichlet boundary conditions. Here, we demonstrate a test example for Neumann boundary conditions in a diffusion-reaction model, namely, the Brusselator model \cite{Jiwari2017}, which was developed by the Brussels school of Prigogine to model the behavior of non-linear oscillators in a chemical reaction system. The model reaction term reads
\begin{align}
    \mathbf{f(u)} = \begin{bmatrix}
        -(k_1+1)u_1 + u_1^2u_2 + k_2 \\
         k_1u_1 - u_1^2u_2
    \end{bmatrix}.
\label{eq:br_f}
\end{align}
Using $\mathbf{D}=[10^{-4},10^{-4}]^T$, $k_1 = 3$ and $k_2 = 1$, with initial conditions
\begin{alignat}{3}
u_1(x,0) = \frac{1}{2}, \qquad && u_2(x,0) = 1+5x, \qquad && x \in (0,1), \\
\frac{\partial u_1}{\partial x}(0,t) = \frac{\partial u_1}{\partial x}(1,t) = 0,  \qquad && \frac{\partial u_2}{\partial x}(0,t) = \frac{\partial u_2}{\partial x}(1,t) = 0, \qquad && t \in [0,T],
\label{eq:br_param}
\end{alignat}
where time $t$ extends up to $T=400$ on $dt = 0.5$.

\Cref{fig:fig5}b shows how a chemical pulse can be spontaneously created, which continually travels leftwards in time, creating traveling waves that appear as striped patterns in time despite low spatial resolutions. 

\section{Applications to the Navier-Stokes Equations}  \label{sec:sec5}
The Navier-Stokes equations are a set of non-linear partial differential equations that describes the motion of fluids across continuum length scales. There are several studies aimed at applying quantum algorithms to computational fluid dynamics (see review \cite{griffin2019investigations}), ranging from reduction of partial differential equations to ordinary differential equations \cite{Gaitan2020} and quantum solutions of sub-steps of the classical algorithm \cite{Steijl2022,Steijl2018} to the quantum Lattice Boltzmann scheme \cite{Budinski2021}.

Here, we look into the potential use of variational quantum algorithms to evolve the fluid momentum equations in time. Consider the incompressible Navier-Stokes equations in non-dimensional form
\begin{align}
    \frac{\partial \mathbf{u}}{\partial t} + \mathbf{u}\cdot\nabla \mathbf{u} &= - \nabla p + \frac{1}{\Rey}\nabla^{2}\mathbf{u}, \label{eq:nse1}\\
    \nabla\cdot\mathbf{u}&=0, \label{eq:nse2}
\end{align}
where $\mathbf{u}$ is the velocity vector and $p$ is the fluid pressure. The ratio $\Rey=U_cL_c/\nu$ is the Reynolds number, where $U_c$ is the characteristic flow velocity across a characteristic length-scale $L_c$ and $\nu$ is the fluid kinematic viscosity.  

Unlike other temporal evolution equations, the incompressible Navier-Stokes equations cannot be time-marched directly as the resultant velocities do not satisfy the continuity constraint \eqref{eq:nse2}, and hence are not divergence-free. To resolve this, the projection method \cite{Chorin1968}, also known as the predictor-corrector or fractional step method, separates the solution time-step into velocity and pressure sub-steps, also known as the \textit{predictor} and \textit{corrector} steps.

\subsection{Projection method}
\subsubsection{Predictor step}
The predictor step first approximates an intermediate velocity $\mathbf{u^{*}}$ by solving the fluid momentum equation \eqref{eq:nse1} in the absence of pressure, i.e.~the Burgers' equations \cite{Oz2022}, of the form
\begin{align}
    \left(1-\frac{\Delta t}{\Rey}\nabla^2\right)\mathbf{u^*}=\left(1-\Delta t\mathbf{u^k}\nabla\cdot\ \right)\mathbf{u^k}.
\end{align}
Through a semi-implicit scheme, the viscous terms are handled implicitly using the variational quantum evolution equation solver and the non-linear inertial terms explicitly as source terms using classical computation. For quantum algorithms for non-linear problems, the reader is referred to separate works on quantum ordinary differential equation solvers \cite{Oz2022,Gaitan2020}, Carleman linearization \cite{Liu2021} and a variational quantum nonlinear processing unit (QNPU) \cite{lubasch2020variational}.

On a two-dimensional domain with Dirichlet boundary conditions, the tensor $\mathbf{A_u}=[A_u,A_v]^T$ and source state $\mathbf{b_u}=[b_u,b_v]^T$ can be decomposed as
\begin{align}
    \begin{gathered}
{\mathbf{A_u}} = I^{\otimes n} + \delta_{x} \mathbf{A_{x, D}} + \delta_{y} \mathbf{A_{y, D}}, \\
\mathbf{\ket{b^k_u}} = (1 - \Delta t \mathbf{F^k}) \mathbf{\ket{u^k}} + \delta_{x}\mathbf{\ket{u_{x,D}^{k+1}}} + \delta_{y}\mathbf{\ket{u_{y,D}^{k+1}}},
\label{eq:decomposed_predictor}
\end{gathered}
\end{align}
where $\delta_x := \Delta t/(\Rey\Delta x^{2})$ and $\delta_y:= \Delta t/(\Rey\Delta y^{2})$. $F^k = D^k_u B_{x,D} + D^k_v B_{y,D}$ is an operator which approximates the non-linear inertial term, where $\mathbf{D^k_{u}}$ are diagonal matrices with velocity vectors $\mathbf{\ket{u^k}}$ along the diagonals and $B$ is a divergence matrix discretized through center differencing, for instance in the $x$ direction, as
\begin{align}
    B_{x,\beta} = \frac{1}{2\Delta x} 
    \begin{bmatrix}
        \alpha_\beta & 1 & 0 & & & \cdots &  & 0 \\
        -1 & 0 & 1 & 0 & & \cdots  & & 0 \\
        0 & -1 & 0 & 1 & & \cdots  & & 0 \\
        \vdots &  &  & \ddots  & & & & \vdots \\
        0 & & \cdots & & 0 & -1 & 0 & 1  \\
        0 & & \cdots & & & 0 & -1 & \alpha_\beta
    \end{bmatrix}.
\end{align}
where $\beta \in \{D,N\}$ refers to either Dirichlet (D) or Neumann (N) boundary condition. Here, $\alpha_D=0$ and $\alpha_N=-1$. 

\subsubsection{Corrector step}
The second corrector step solves for the velocity $\mathbf{u^{k+1}}$ by correcting the intermediate velocities $\mathbf{u^*}$ using the pressure gradient as a Lagrange multiplier to enforce continuity. Applying divergence to the correction equations yields the \textit{pressure Poisson equation} for the pressure field at half-step
\begin{align}
-\nabla^2 p^{k+1} = -\frac{1}{\Delta t}\nabla\cdot{\mathbf{u^*}},
\end{align}
which can be solved implicitly in two dimensions $(x,y)$ via the following decomposition: 
\begin{align}
\begin{gathered}
A_{p} = \frac{\Delta t}{\Delta x^2}\left(A_{x,N} + \frac{1}{2}I_0^{\otimes n}\right) + \frac{\Delta t}{\Delta y^2}\left(A_{y,N} + \frac{1}{2}I_0^{\otimes n}\right), \\
\ket{b^k_p} = -(B_{x,D}\ket{u^*} + B_{y,D}\ket{v^*}).
\label{eq:decomposed_corrector}
\end{gathered}
\end{align}
Note the addition of a simple Hermitian $I_0 = \kbra{0}{0}$ to the pressure matrix $A_p$, which would otherwise be singular (corank 1) under Neumann boundary conditions for the pressure field. 

With the new pressure $p^{k+1}$, the velocities are updated at the $k+1$ time-step as
\begin{align}
\mathbf{u^{k+1}} = \mathbf{u^*} - \Delta t \mathbf{B_N} \ket{p^{k+1}},
\end{align}
where $\mathbf{B_N} = [B_{x,N}, B_{y,N}]^T$ are the gradient operators.

\subsection{Implementation}
Overall, the variational quantum solver for Navier-Stokes equations using the projection method (see Algorithm \ref{algo03}) involves two sequential steps, the first requiring a number of Algorithm \ref{algo01} iterations equal to the number of velocity components, and the second for the pressure Poisson step. For two-dimensional flows, the number of velocity components to be solved can be effectively reduced by one through the vorticity stream-function formulation (Appendix \ref{appendix:vorticity_streamfunc}). In computational fluid dynamics, these implicit systems of linear equations are often the most computationally expensive parts to solve in classical algorithms, providing incentives for potential speedup via quantum computing \cite{griffin2019investigations,Steijl2018}. 

\begin{minipage}{0.95\linewidth}
\noindent
\begin{algorithm}[H]
	\caption{Variational Quantum Navier-Stokes Equation Solver (Projection Method)}
	\label{algo03}
	\begin{algorithmic}[3]
	    \State {Initialize} $\mathbf{\ket{\psi(\theta)}}$ for $\{\mathbf{u},p\}$
		\For {$k \in [0,n_t)$}
			\State {Encode} $\mathbf{\ket{\hat{b}^{k}_u}} \gets \mathbf{\ket{u^k}}$	(Eq.~\ref{eq:decomposed_predictor})	 
			\While {$\varepsilon_\mathbf{u} > \varepsilon_{\tol}$}
			    \State {Evaluate} $E(\mathbf{\theta^k_u})$ on a quantum computer for $\mathbf{A_u}$ and $\mathbf{\ket{\hat{b}^k_u}}$
				\State Update $\mathbf{\theta^k_u}$ using classical optimization 
			\EndWhile
			\State {Update} $\mathbf{\ket{u}^*} \gets \mathbf{r(\theta^k_u}) \mathbf{\ket{\psi(\theta^k_u})}$
			\State {Encode} $\ket{\hat{b}^{k}_p} \gets \mathbf{\ket{u}^{*}}$ (Eq.~\ref{eq:decomposed_corrector})
			\While {$\varepsilon_p > \varepsilon_{\tol}$}
			    \State {Evaluate} $E(\theta^k_p)$ on a quantum computer for $A_p$ and $\ket{\hat{b}^k_p}$
				\State Update $\theta^k_p$ using classical optimization 
			\EndWhile
			\State {Update} $\ket{p^{k+1}} \gets r(\theta^k_p)\ket{\psi(\theta^k_p)}$
            \State {Update} $\mathbf{\ket{u^{k+1}}} \gets \mathbf{\ket{u^*}},\ket{p^{k+1}}$
		\EndFor
		\State {Return} $\mathbf{\ket{u}},\ket{p}$
	\end{algorithmic} 
\end{algorithm}
\end{minipage}
\vspace{\baselineskip}\linebreak

\subsection{2D Cavity Flow}
The lid-driven cavity flow is a standard benchmark for testing incompressible Navier-Stokes equations \cite{Erturk2005}. Consider a two-dimensional square domain $\Omega =(0,L)\times (0, L) \subset \mathbb R^2$ with only one wall sliding tangentially at a constant velocity. For simplicity, we employ a fixed collocated grid, instead of a staggered grid which helps avoid spurious pressure oscillations but at the cost of increased mesh and discretization complexity. No-slip boundary conditions apply on all walls, so that zero velocity applies on all wall boundaries except one moving at $u(x,0)=1$.

\begin{figure}
\centering
  \includegraphics[width=0.9\linewidth]
  {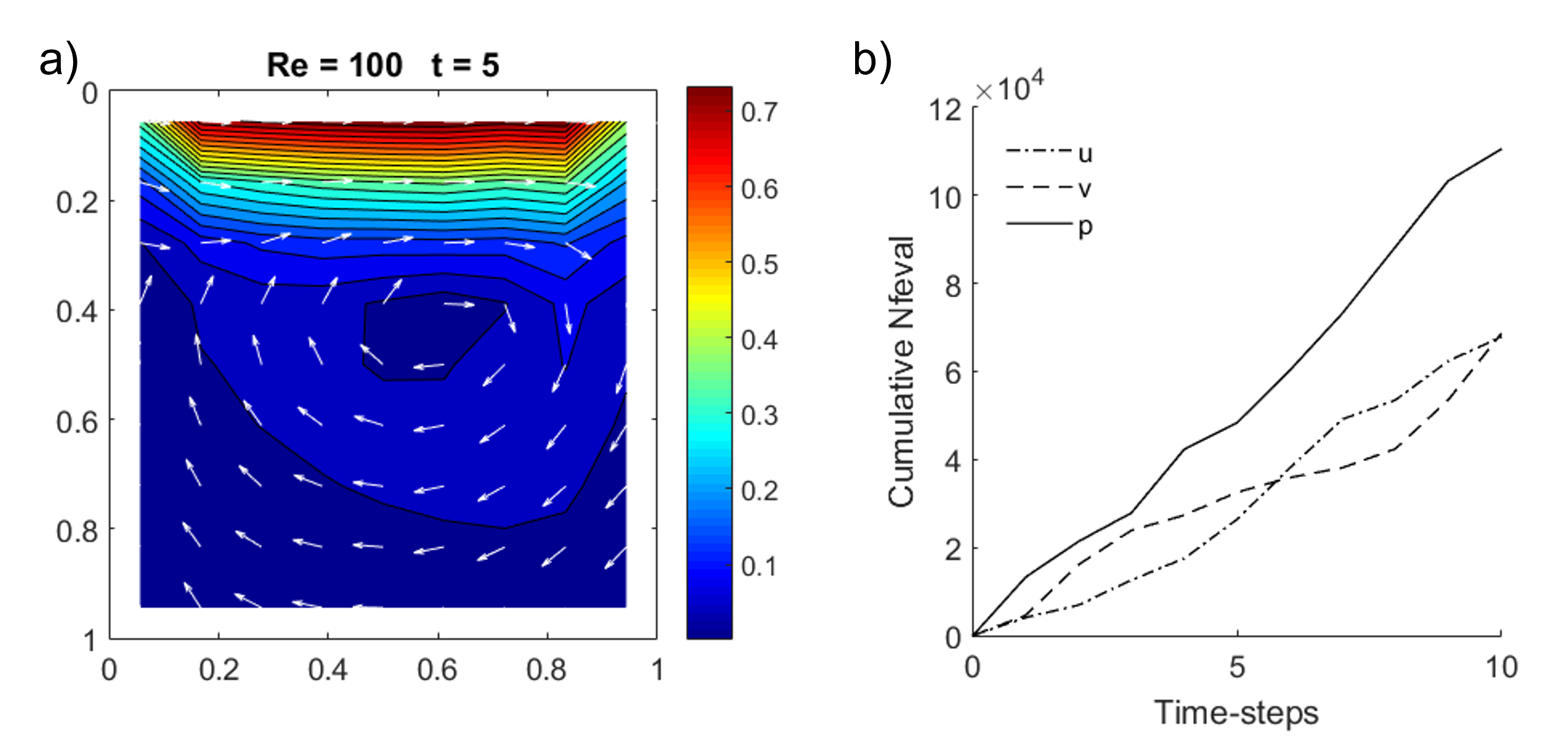}
\caption{(a) 2D lid-driven cavity flow ($\Rey = 100$) on a $2^n=8\times8$ grid at $\Delta t=0.5$ up to $T = 5$ and upper boundary sliding in the $x$ direction at $u(x,0)=1$. Color map indicates velocity magnitude with normalized velocity quivers in white indicating direction of flow. (b) Plots of cumulative number of function evaluations vs.~time for intermediate velocities ($u$,$v$) and pressure ($p$).}
\label{fig:fig6}
\end{figure}

Figure \ref{fig:fig6}a shows a snapshot of a test case conducted on a $2^n=8\times8$ grid at $\Delta t=0.5$ up to $T = 5$, with the central vortex shown by normalized velocity quivers in white. In terms of time complexity, we note that the pressure correction step requires a greater number of function evaluations for convergence compared to an implicit velocity step (Fig.~\ref{fig:fig6}b). This is due to the additional quantum circuits for evaluating the $H_4$ Hamiltonians (\ref{eq:A2a_decomposition}, \ref{eq:A2b_decomposition}) for Neumann boundary conditions and one for specifying the reference pressure (\ref{eq:decomposed_corrector}), leading to a total of 9 evaluation terms compared to 6, a ratio which corroborates with the apparent $\sim50\%$ increase in function evaluations shown in Fig.~\ref{fig:fig6}b.

While not directly comparable to classical computational fluid dynamics in numerical accuracy, this exercise, nevertheless, roadmaps potential applications of the variational quantum method towards more complicated flow problems \cite{Steijl2018}.

\section{Conclusion}
In this study, we proposed a variational quantum solver for evolution equations which include a Laplacian operator to be solved implicitly. For short time-steps $\Delta t$, the use of initial parameter sets encoded from prior solution vectors results in faster convergence compared to random re-initialization. The overall time complexity scales with the ansatz volume $\mathcal{O}(nl)$ for gradient estimation and with the number of time-steps $\mathcal{O}(n_t)$ for temporal discretization. Our proposed algorithm extends naturally to higher-order time-stepping and higher dimensions. For evolution equations with non-trivial source terms, the semi-implicit scheme can be applied, where non-linear source terms are handled explicitly. Using statevector simulations, we demonstrated that variational quantum algorithms can be useful in solving popular partial differential equations, including the reaction-diffusion and the incompressible Navier-Stokes equations. 

The present work aims at bridging the gap between variational quantum algorithms and practical applications. Our work has assumed that the state preparations, unitary transformations and measurements are implemented perfectly, and does not consider the effects of quantum noise from actual hardware or any potential amplification from iterative time-stepping. In our implementation, we only considered the hardware-efficient ansatz with $R_y$ rotation gates and controlled-NOT entanglers, and thus leave open the question about the performance of other ansatzes. Future work can include noise mitigation \cite{li2017efficient, temme2018error, endo2018practical
}, ansatz architecture, non-linear algorithms and cost-efficient encoding. 

\section*{Acknowledgements}

We thank Maria Schuld for helpful comments on amplitude embedding. This work was supported in part by the Agency for Science, Technology and Research (A*STAR) under Grant No.~C210917001. DEK acknowledges funding support from the National Research Foundation, Singapore, through Grant NRF2021-QEP2-02-P03.

\appendix
\numberwithin{equation}{section}

\section{Fully implicit scheme for linear Hermitian source term } \label{appendix:implicit}
For a reaction-diffusion system of equations with non-trivial source terms \ref{eq:rd_initial}, time-stepping could be rendered fully implicit if the source terms can be expressed as a linear Hermitian matrix with constant coefficients. Consider a two-component 1D chemical reaction with source terms of the form
\begin{align}
    \mathbf{f(u)} = \begin{bmatrix}
        k_{11} & k_{12} \\
        k_{21} & k_{22}
    \end{bmatrix} \otimes \mathcal{I} 
    \begin{bmatrix}
        u_1 \\
        u_2
    \end{bmatrix},
\end{align}
where $k_{ij}$ ($i,j \in\{1,2\}$) are kinetic rate constants that are components of the linear Hermitian matrix $K = K^T \in \mathbb R^{N\times N}$, whose off-diagonal elements are equal, i.e.~$k_{12}=k_{21}$; $\mathcal{I}$ is the identity matrix of size $N\times N$ and $[u_1, u_2]^T$ is a concentration vector of length $2N$. This problem requires $n+1$ qubits, where $n=\log_2N$. Following the implicit Euler scheme (Eq.~\ref{eq:IE}), we set up a $2N\times2N$ coefficient matrix, which decomposes as 
\begin{align}
    A = I^{\otimes n+1} + \delta_x I\otimes A_{x,\beta} - \Delta t\left(k_{11}I_0 + k_{22}I_1 + k_{12}X \right) \otimes I^{\otimes n},
\end{align}
where $A_{x,\beta}$ is the discretized $N \times N$ coefficient matrix for a single component (Eq.~\ref{eq:A_decomposition}) containing up to four Hamiltonian terms $H_{1-4}$. Note the last three additional Hamiltonian terms contributed by the source term.

\section{Vorticity stream-function formulation}
\label{appendix:vorticity_streamfunc}
For two-dimensional incompressible flows, the vorticity stream-function formulation can be used to eliminate the pressure as a dependent variable, such that
\begin{align}
    \frac{\partial \omega}{\partial t} + u \frac{\partial\omega}{\partial x} + v \frac{\partial\omega}{\partial y} &= \frac{1}{\Rey} \left( \frac{\partial^2\omega}{\partial x^2} + \frac{\partial^2\omega}{\partial x^2} \right), \label{eq:vse1}\\
    \frac{\partial^2\psi}{\partial x^2} + \frac{\partial^2\psi}{\partial x^2} &= -\omega, \label{eq:vse2}
\end{align}
where $\omega = \partial v/\partial x - \partial u/\partial y$ is the flow vorticity and the stream-function $\psi$ satisfies $u=-\partial\psi/\partial y$ and $v=\partial\psi/\partial x$. It follows that the re-formulation $\{u,v,p\} \rightarrow \{\omega,\psi\}$ can simplify the variational quantum algorithm \ref{algo03} by reducing the number of velocity components by one, at the cost of specifying stream-function values along the domain boundaries. 

\bibliographystyle{unsrt}
\bibliography{bib}

\end{document}

%% file: setup.tex
\usepackage[letterpaper,top=2cm,bottom=2cm,left=2cm,right=2cm,marginparwidth=2cm]{geometry}
\usepackage[font=small,margin=2cm]{caption}
\usepackage{authblk}
\usepackage[colorlinks,citecolor=BlueViolet,urlcolor=blue,bookmarks=false,hypertexnames=true]{hyperref}
\usepackage{cite}
\usepackage{graphicx}
\usepackage{algorithm}
\usepackage{algpseudocode}
\usepackage[dvipsnames]{xcolor}
\usepackage{orcidlink}

\usepackage{amsmath}
\usepackage{amssymb}

\usepackage[capitalise]{cleveref}

\usepackage{qcircuit}
\newcommand{\ket}[1]{| #1 \rangle}
\newcommand{\bra}[1]{\langle #1 |}
\newcommand{\kbra}[2]{|#1\rangle\!\langle #2|}
\newcommand{\bket}[2]{\langle #1|#2\rangle}
\DeclareMathOperator{\tr}{tr}
\def\opt{\mathrm{opt}}
\def\eval{\mathrm{eval}}
\def\tol{\mathrm{tol}}
\def\Rey{\mathrm{Re}}

\providecommand{\keywords}[1]
{
\noindent
    {\small
  \textbf{Keywords}: #1}
}

\DeclareMathSymbol{\mh}{\mathord}{operators}{`\-}
\usepackage{appendix}